\documentclass[
 reprint,
 amsmath,amssymb,
 aps,
 prx
]{revtex4-1}

\usepackage{graphicx}
\usepackage{natbib}
\usepackage{textcomp}
\usepackage{dcolumn}
\usepackage{bm}
\usepackage{latexsym}
\usepackage{psfrag}
\usepackage{xcolor}
\usepackage{color}
\usepackage{epstopdf}
\usepackage{amsmath}
\usepackage{wasysym}
\usepackage{comment}

\newcommand\Bo{\mbox{\textit{Bo}}}  

\begin{document}

\preprint{APS/123-QED}

\title[Highly focused supersonic microjets]{Highly focused supersonic microjets}% Force line breaks with \\
%\thanks{A footnote to the article title}%

\author{Yoshiyuki Tagawa$^{1}$}
 \email{y.tagawa@utwente.nl}
 \author{Nikolai Oudalov$^{1}$}%
% \affiliation{$^{1}$Physics of Fluids, University of Twente\\
%PO Box 217, 7500 AE, Enschede, The Netherlands 
% }%Lines break automatically or can be forced with \\
\author{Claas Willem Visser$^{1}$}
% \affiliation{Physics of Fluids, University of Twente\\
%PO Box 217, 7500 AE, Enschede, The Netherlands 
% }%Lines break automatically or can be forced with \\
\author{Ivo R. Peters$^{1}$}
% \affiliation{Physics of Fluids, University of Twente\\
%PO Box 217, 7500 AE, Enschede, The Netherlands 
% }%Lines break automatically or can be forced with \\
\author{Devaraj van der Meer$^{1}$}
% \affiliation{Physics of Fluids, University of Twente\\
%PO Box 217, 7500 AE, Enschede, The Netherlands 
% }%Lines break automatically or can be forced with \\
\author{Chao Sun$^{1}$}
% \affiliation{Physics of Fluids, University of Twente\\
%PO Box 217, 7500 AE, Enschede, The Netherlands 
% }%Lines break automatically or can be forced with \\
\author{ Andrea Prosperetti$^{1, 2}$}
% \affiliation{Physics of Fluids, University of Twente\\
%PO Box 217, 7500 AE, Enschede, The Netherlands 
% }%Lines break automatically or can be forced with \\
% \altaffiliation{Department of Mechanical Engineering, Johns Hopkins University, Baltimore, MD 21218, USA}
\author{Detlef Lohse$^{1}$}
% \affiliation{Physics of Fluids, University of Twente\\
%PO Box 217, 7500 AE, Enschede, The Netherlands 
% }%Lines break automatically or can be forced with \\
\affiliation{$^{1}$Physics of Fluids group, MESA+ Institute $\&$ Faculty of Science and Technology,
University of Twente, P.O. Box 217, 7500 AE Enschede, The Netherlands\\ 
 $^{2}$Department of Mechanical Engineering, Johns Hopkins University, Baltimore, MD 21218, USA}

\date{\today}
\begin{abstract}

%The development of needle-free drug injection systems is of paramount importance to the global fight against the spread of disease. Successful application of these systems requires highly-focused microjets with ultra-high velocities (more than 200 m/s) and great controllability.
The paper describes the production of thin, focused microjets with velocities
up to 850 m/s by the rapid vaporization of a small mass of liquid in an open
liquid-filled capillary. 
The vaporization is caused by the absorption of a low-energy laser pulse. 
A likely explanation of the observed phenomenon is based on the impingement of the shock wave
caused by the nearly-instantaneous vaporization on the free surface of the liquid.
%The generation of thin, high velocity focused microjets with velocities up to 850\,m/s is described.
%The microjets are produced by focusing a laser pulse in a liquid-filled glass microcapillary. 
%A small mass of liquid is instantaneously vaporized, which causes the overpressure responsible for the generation of the high-speed microjet. 
%A likely explanation of the process based on the impingement on the liquid free surface of the shock wave caused by the rapid vaporization is proposed.
%We demonstrate a novel method of creating microjets with a very sharp geometry and controlled velocities even for supersonic speeds
An experimental study of the dependence of the jet velocity on several parameters is conducted, and a semi-empirical relation for its prediction is developed.
%is , discuss the physical background of these results, 
%and arrive at scaling relations for the jet velocity as function of the various parameters. 
The coherence of the jets, their high velocity and good reproducibility and controllability are unique features of the system described.
A possible application is to the development of needle-free drug injection systems which are of great importance for global health care.% and requires highly-focused microjets with ultra-high velocities (more than 200 m/s) and good reproducibility.

\begin{description}
%\item[Usage]
%Secondary publications and information retrieval purposes.
\item[Subject Areas]
Fluid dynamics
%\pacs{Flows and jets through nozzles} 47.60.Kz 
%\pacs{Micro- and nano- scale flow phenomena} 47.61.-k 
%\pacs{Supersonic and hypersonic flows} 47.40.Ki	

%May be entered using the \verb+\pacs{#1}+ command.
%\item[Structure]
%You may use the \texttt{description} environment to structure your abstract;
%use the optional argument of the \verb+\item+ command to give the category of each item. 
\end{description}
\end{abstract}

\pacs{Valid PACS appear here}% PACS, the Physics and Astronomy
                             % Classification Scheme.
%\keywords{Suggested keywords}%Use showkeys class option if keyword
                              %display desired
\maketitle

\section{Introduction}
\label{sec:intro}

Transient liquid jets produced by flow focusing have been studied from various perspectives due to  fundamental interest and their widespread occurence (see e.g. the review article by \citet{Eggers2008}). 
%The way in which these jets are obtained show great variation, such as tubular jets \cite{Bergmann2008, Antkowiak2007, Lorenceau2002}, jets caused by bubbles bursting at free surfaces \cite{Duchemin2002, , jets from breaking waves or , and those from collapsing voids 
Common instances include bubbles bursting at a liquid surface \cite{Duchemin2002,LonguetHiggins1983}, high amplitude Faraday waves \cite{LonguetHiggins2001a, LonguetHiggins2001b, LonguetHiggins2001c, Zeff2000, Goodridge1999}, jets generated by the impulsive acceleration of a liquid surface \cite{Antkowiak2007}, shaped charges, and others.\\
%The principle is also used for the so-called tubular jet, where a meniscus is accelerated into an evacuated tube forming a jet \cite{Lorenceau2002, Bergmann2008}, and the jets developing in a suddenly decelerating falling partially liquid-filled tube \cite{Antkowiak2007}. % quantitatively analyzed jets produced by dropping a partially filled test tube on a hard 
%}%, to quote just a few examples among many [432]. 
% include ink-jet printing \cite[]{deJong2006, Wijshoff2010, VanHoeve2010}, ultrasound surface cleaning  \cite[]{Ohl2006,Dijkink2008,Zijlstra2009}, the damage done by imploding cavitation bubbles \cite[]{Blake1997}. 
%However, all these jets either do not have particular high velocities (i.e. they are at most of the order of 10 m/s) or they are rather uncontrolled.
In all these cases, the jet is produced when a  liquid surface concave towards the gas is impulsively accelerated.
The resulting ``kinematic focusing" of the liquid, which converges towards the center of the curvature, is responsible for the peculiarly high-speed jets generated in this way.\\
In shaped charges this convergence is caused by a detonation wave investing the free surface. The early history of these devices and the principle on which they operate is described by \citet{Birkhoff1948}, followed by many works, e.g., \citet{Curtis1994, Petit2005} in more recent years. 
The focusing, however, can also be caused by geometrical effects simply due to the concave shape of the interface without reliance on liquid compressibility (see e.g. \cite[]{LonguetHiggins1995, Bergmann2006, Bergmann2009, Thoroddsen2007, Gekle2009, Gekle2010}).\\
In this study we describe yet another way for the generation of such a jet. We use a microtube in which the impulsive acceleration of the interface is due to the abrupt vaporization of a small mass of liquid caused by the absorption of a laser pulse. \\
Bubble generation by the absorption of a focused laser pulse in the bulk of a
liquid volume has been an established experimental technique for a long time
(see e.g.~\cite{Lauterborn1999,Vogel2003}). More
recently, the same technique has been used to generate bubbles in the
proximity of the free surface of a liquid~\cite{Apitz2005,Obreschkow2006,Robert2007,Thoroddsen2009}.
In this case, strong jets are often observed. For example, \citet{Thoroddsen2009}, who focused a laser pulse into a drop resting on a glass surface, report jet maximum velocities of 250 m/s produced by small bubbles near the free surface, but without the ability to control the number, size and location of these bubbles.
%\citet{Thoroddsen2009} used a focused laser pulse inside the drop sits on a glass surface, observing the microjets from the submerged small bubbles at the top of the droplet. The small bubbles have been generated by focusing laser pulses in liquids for decades.
%The maximum velocity of the water microjet is 250 m/s.
%\citet{Robert2007} creates laser-induced vapor cavities inside axisymmetric free-falling liquid water jets and observed microjets in small cavitation bubbles inside the jets.
%Although the idea of creating the jet is similar, the geometry of the present system is quite different. Furthermore, they have less control over the number, sizes or location of the small bubbles. % but mainly reported observations.
\\
% the sudden appearance of the vapor bubble produced by a focused laser pulse near the free surface containing in the capillary tube with the diameter of the order of 100 $\mu$m.}
%Unlike our work, these authors  do not pursue controllability of the high velocity jets.surface. 
%\textcolor{red}{The shaped-charge effect introduced here is the effect when the liquid surface obtains normal velocities concentrated in the void (i.e., a concave surface set into motion) the liquid surface accelerates due to the focusing of the liquid velocity, creating a jet.}
%\textcolor{red}{The uniqueness of the present jet comes from using a shock wave, created from the
%What happens is that the shock wave, impinging on the free surface, imparts to the liquid a velocity directed normal to the free surface and therefore, because of the curvature, 
%directed toward the tube axis. It is this ``kinematic focusing" that causes the large velocity of the jet. The present microjet, which is created by this mechanism, has never been reported before. 
%}
%To create our supersonic microjets, 
%In this study, we present a unique method for the creation of high-speed focused microjets.
% created% by a laser-induced vapor bubble.% (figure~\ref{fig:iLIF}).
In the present work we also use a laser pulse to create a bubble near a free
surface, but the geometry of the system and condition of the experiment
are quite different from those of earlier workers. Furthermore, the microjet
that we describe has extraordinary characteristics: it is created under
controlled condition with good reproducibility; its maximum velocity reaches
850 m/s, namely a Mach number in air approaching 3; its diameter is typically one order of magnitude smaller than the diameter
of the capillary tube; its generation requires moderate
laser energies of the order of 10 $\mu$J, comparable to the
typical energy of a laser pointer.\\
%The microjet that we describe has extraordinary characteristics:
%\textcolor{red}{it is created under controlled condition with good reproducibility;}
%its maximum velocity reaches 850 m/s, a Mach number greater than 2;
%its diameter is typically one order of magnitude smaller than the diameter of the capillary tube; % and is decoupled from it.
%its generation requires moderate laser energies of the order of 10 $\mu$J, comparable to the typical energy of a laser pointer. \\
%This is much faster than the velocities which can be achieved with conventional techniques.
% The required energy thus is far smaller than for the conventional  method proposed by \citet{Han2010}. 
%The shape of present microjet is highly-focused and axisymmetric -- another  feature that is absent from conventional jetting devices. Note that  }%The problem of clogging has rarely occurred when using this method.}
A potential particularly attractive application of the present supersonic microjets is for {\it needle-free drug injection} \cite[]{Mitragotri2006}.
For this application, in which a liquid solution containing a drug is forced to penetrate human or animal tissue through the skin, ultra-high velocities (greater than 200 m/s), fine scales (down to 30 $\mu$m), and good reproducibility and controllability are essential.  \\
%For this last purpose, the important requirements for these microjets are:
%ultra-high speed  good controllability, 
%The aim of these devices is to jet 
%   \cite[]{Mitragotri2006}. 
The development of needle-free drug injection systems is of paramount importance in the global fight against the spread of disease as these systems help preventing contamination by needles.
Several methods of microjet generation suitable for this application have been proposed (\cite{Han2010,Menezes2009}), but they all have several shortcomings: the production of each jet requires a siginificant amount of energy, the resulting shape of the ejected liquid mass tends to be diffuse rather than focused, the small size-nozzles are easily clogged and others. %which affects.
%In the present study, we describe a novel method for the generation of even faster and thinner jets and the way in which their velocity can be controlled by the suitable choice of experimental parameters.
%\textcolor{blue}{Despite the efforts mentioned above, crucial questions about our micro-scale jets with high velocities remain.
%How far can we upscale the velocity of the microjet? 
\\
The microjets produced by the method described in this study are free from these problems. 
In particular, they are faster and highly focused. Furthurmore, the avoidance of very thin nozzles makes the systems less susceptible to clogging, thus improving the reliability and controllability.
\\
The experimental setup is presented in \S~\ref{sec:setup}, followed by a description of jet formation and evolution in \S~\ref{sec:results}.
The quantitative dependencies of the jet speed on the various control parameters are discussed in \S ~\ref{sec:parameter_results}, culminating in a scaling law for the jet velocity as a function of the various parameters.
The paper ends with summary and conclusions in \S ~\ref{sec:conclusions}.
%gives rise to well-defined jets with diameters well below 20 $\mu$m and supersonic velocities. 
%The questions we adrress are: 
%How can the mechanisms proposed above be applied on the micro-scale?
%How does the use of a shock wave for jet generation change things?
%Which physical quantities affect the dynamics of the microjet and how to account for the found dependences? 
%How fast can the microjet be? %How well can we control the microjets?

\section{Experimental setup and control parameters} \label{sec:setup}
\begin{figure}
\centerline{\includegraphics[width=.4\textwidth]{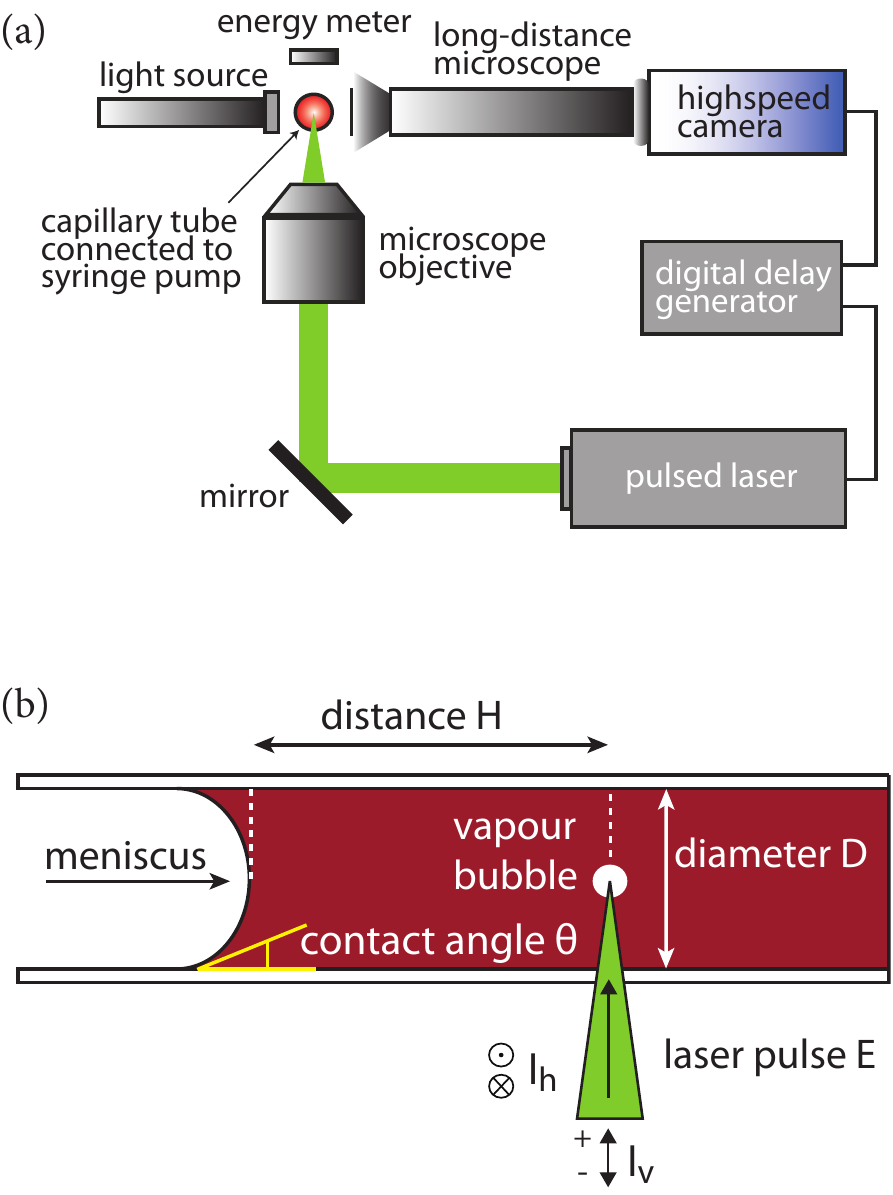}}
\caption{\label{fig:full_setup} (a) A sketch of the experimental setup; the capillary tube is aligned perpendicular to the paper. (b) Definition of experimental parameters with a side view of the capillary in which the microjet formation takes place.}
\end{figure}

Figure~\ref{fig:full_setup}(a) shows a sketch of the whole experimental setup, which is similar to that used by \citet{Sun2009}. One end of the capillary tube is connected to a syringe pump (Model PHD 2000, Harvard Apparatus, USA) containing a water-based red dye. The other end is open to the air. A 532\,nm, 6\,ns laser pulse (100\,mJ Nd:YAG laser, Solo PIV, New Wave, USA) is focused through a $10\times$ microscope objective to a point inside the capillary. The laser energy is monitored by means of an energy meter (Gentec-eo XLE4 or Gentec-eo QE12SP-S-MT-D0, Canada) placed behind the capillary. The energy absorbed by the liquid was calibrated by measuring the difference between the readings of the meter with the glass tube filled with the dyed and clear water. The jet formation was recorded using high-speed cameras with a frame rate of up to $10^6$ fps (HPV-1, Shimadzu Corporation, Japan and SA1.1 and Photron, USA). The minimum inter-frame and exposure times were 1$\,\mu$s and $250\,$ns respectively. This system enabled us to observe the capillary from below using a microscope (Axiovert 40 CFL, Carl Zeiss, Germany). A long-distance microscope (Model K2, Infinity, USA) with a maximum magnification of $12\times$ was connected to the camera in order to capture the jet formation from the side and vary the field of view by adjusting the magnification. 
Illumination for the camera was provided by a fibre lamp (ILP-1, Olympus, Japan) emitting light that passed through the filter protecting the camera. A digital delay generator (Model 555, Berkeley Nucleonics Corporation, USA) was used to synchronise the camera and the laser. Images were analysed with tracking software. 
\\
The experimental parameters are indicated in figure~\ref{fig:full_setup}(b) together with a side view of the capillary. The distance between the meniscus and the laser focus position is $H$, $E$ is the absorbed laser energy, $\theta$ is the initial contact angle of the liquid with the glass capillary with diameter $D$; $l_{v}$ is the distance offset of the laser focus with respect to the capillary axis along a vertical diameter, positive when the focus is beyond the tube axis; $l_{h}$ is the focus offset of the laser perpendicular to the tube axis and the diameter along which $l_v$ is measured. Borosilicate glass capillary tubes (Capillary Tube Supplies Limited, UK) of three different sizes were used in the experiments, with inner diameters $D$ of 50, 200 and 500$\,\mu$m and outer diameters of 80, 220 and 520$\,\mu$m, respectively. The tip of the capillary tube was dipped in a hydrophobic solution (1H, 1H, 2H, 2H-Perfluorooctyltrichlorosilane) to pin the contact line and permit a control the contact angle $\theta$ between $20^\circ$ and $90^\circ$. For uncoated tubes, the initial contact angle was found to be $25^{\circ}\pm{3^\circ}$, which is similar to the data reported by \citet{Sumner2004} for borosilicate glass. The range of the control parameters is summarised in table~I.%\ref{tab:param}.
\\

\begin{table}
\label{tab:param}
\caption{The range of control parameters for the initial contact angle of the liquid with the glass capillary $\theta$, the distance between the meniscus and the laser focus position $H$, the absorbed laser energy $E$, the focus offset of the laser in the vertical plane $l_{v}$, and the focus offset of the laser in the horizontal plane $l_{h}$. The inner diameters of capillary tubes $D$ are 50, 200, and 500$\,\mu$m.}
\begin{center}
 \begin{tabular}{| c | c | c | c | c | c | c | c}
 \hline 
                     &$\theta$      	&$E$($\,\mu$J)&	$H$($\,\mu$m)	& $D$ ($\,\mu$m)	& $l_v$ (mm)	& $l_h$(mm)\\ \hline 
  Lower limit  & 20$^\circ$			&   19  		&  $200$		& 50 				& -1 			&0	\\ 
  Upper limit  & 90$^\circ$			& 880 		&  $2500$ 		& 500  			&  2.5		&0.1\\ \hline 
    \end{tabular}
    \end{center} 
\end{table}

\section{Jet formation and evolution}\label{sec:results}
\begin{figure*}
\centerline{\includegraphics[width=.7\textwidth]{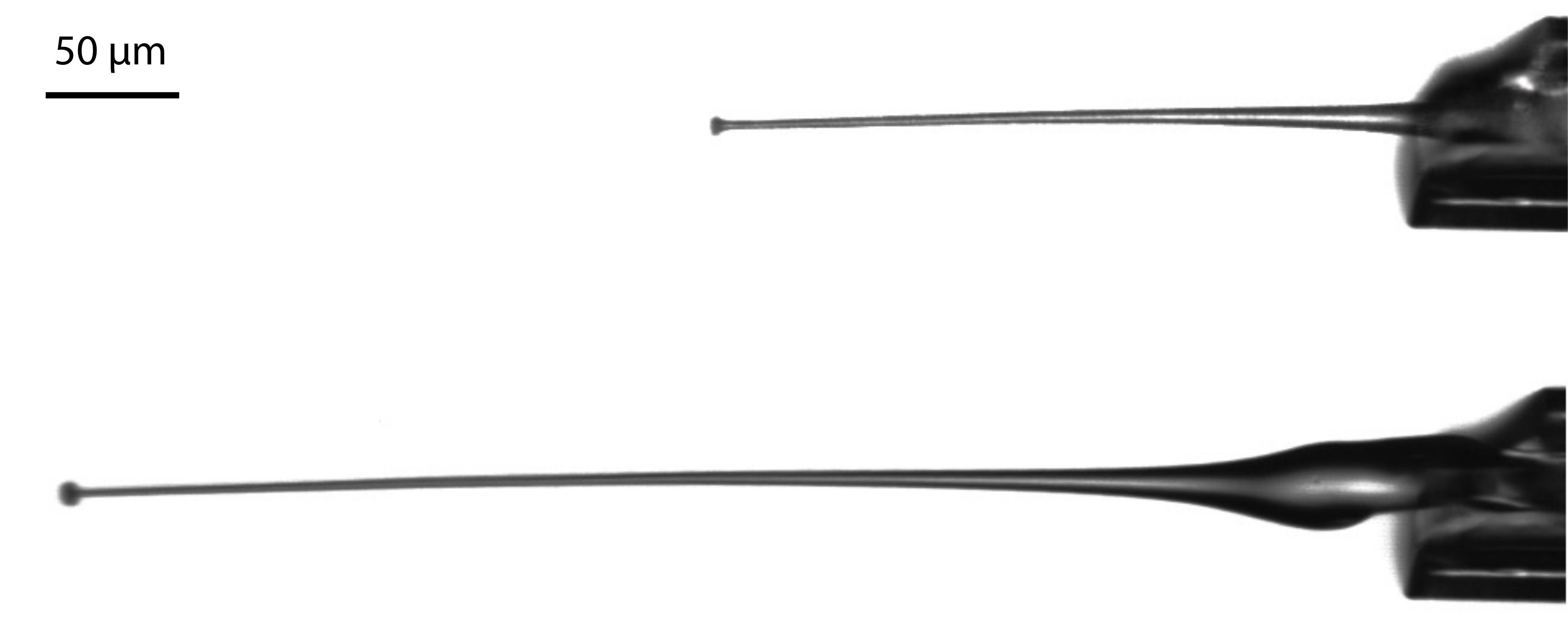}}
\caption{\label{fig:iLIF} Images of a supersonic microjet generated in a 50$\,\mu$m capillary tube. The capillary is visible on the right side, the jet tip is shown on the left. The jet travels from right to left with a speed of 490\,m/s. Time between images is 500\,ns.}
\end{figure*}

\begin{figure}
\centerline{\includegraphics[width=.5\textwidth]{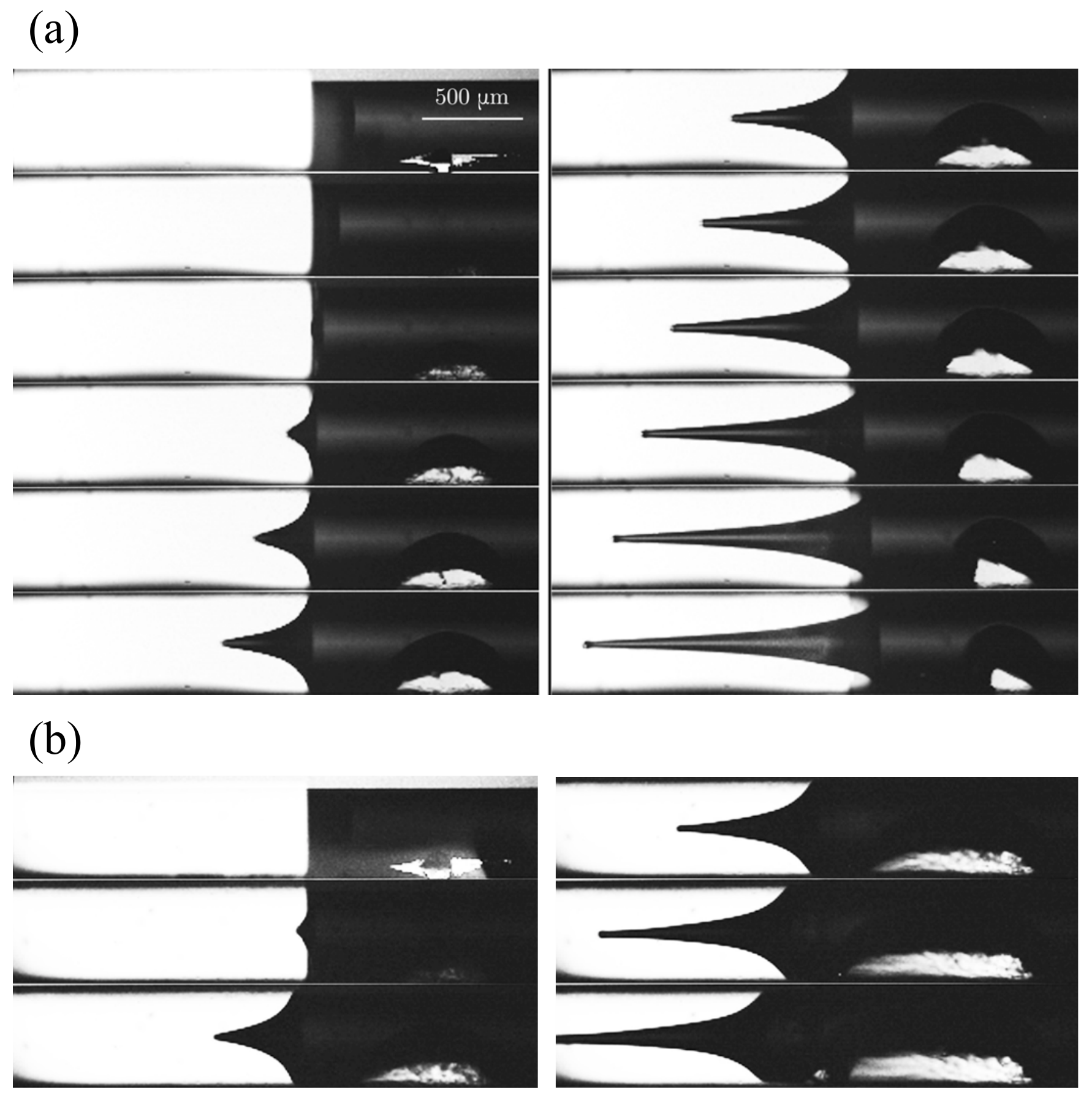}} %fig3.pdf
\caption{\label{fig:jet_evolution} Bubble growth and jet evolution after focusing the laser on the axis of a 500$\,\mu$m capillary tube with (a) $E$ = 365 $\mu$J and $H$ = 600 $\mu$m  and (b) $E$ = 650 $\mu$J and $H$ = 600 $\mu$m.The first image shows the tube at the moment the laser is fired. The subsequent images are taken 7$\,\mu$s apart. (The movie is available as supplemental material [figure3.mov] )}
\end{figure}

\begin{figure}
\centerline{\includegraphics[width=.4\textwidth]{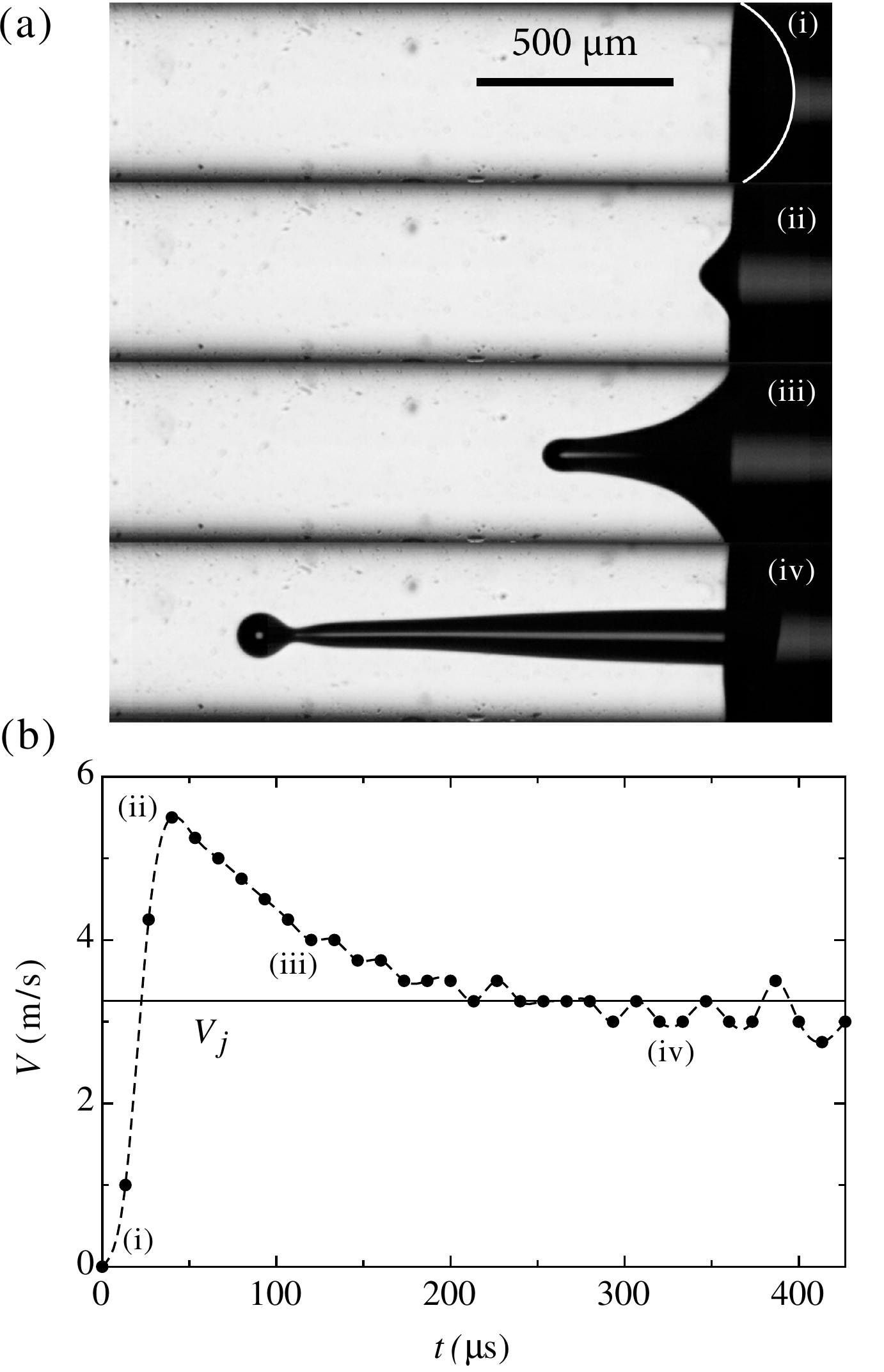}}
\caption{\label{fig:time_evolution} (a) Jet evolution after focusing the laser in a 500$\,\mu$m capillary tube. The first image shows the tube at the moment that the laser is fired. The subsequent images are taken at later stages, each corresponding to a point in time shown in the adjacent figure. (The movie is available as supplemental material  [figure4a.mov].) (b) Velocity of the jet tip in a 500$\,\mu$m tube as a function of time after the laser is fired.}
\end{figure}

An example of a supersonic jet visualized using the technique of dual-flash illumination by laser-induced fluorescence \cite{VanderBos2011a} is shown in figure ~\ref{fig:iLIF}. The capillary is on the right, the jet tip is on the left, and the jet travels from right to left with a speed of 490\,m/s. The tip of this jet has grown into an almost spherical mass which will eventually detach as a droplet. At the other end, the jet is thicker and slower.  
\\
Figures~\ref{fig:jet_evolution}(a) and (b) show successive frames taken with the Shimadzu camera of the jet produced in a 500 $\mu$m tube by using laser energies of 365 $\mu$J and 650 $\mu$J.
The bubble growth is larger in the second case and the motion of the liquid column in the tube is more evident. 
The use of this larger capillary produces relatively slow jets with asymptotic velocities of $\sim$21 m/s and 55 m/s, respectively, which facilitates the visualization.
The open end of the tube is on the left; the other end is connected to the syringe pump as explained before. 
The bubble is generated near the lower wall of the capillary and mainly expands in the direction of the open end.
In spite of this off-center position and asymmetric growth, the jet maintains an axisymmetric shape with a sharp tip.
Its diameter is always much smaller than that of the capillary.
In this instance, it is $\sim$50 $\mu$m, about 10 times smaller than the capillary.
The beginning of the instability which eventually leads to droplet pinch off can be discerned. 
We have always found that, unless the bubble is generated less than one diameter away from the free surface ($H \le D$), the asymmetry of its growth does not affect the axial symmetry of the jet. 
This observation supports the hypothesis that the cause of the thin jet is the sudden onset of a high pressure rather than the bubble expansion \textit{per se}.
\\
Figure~\ref{fig:time_evolution}(a) shows another typical sequence of jet evolution. The corresponding history of the jet tip velocity is shown in Figure~\ref{fig:time_evolution}(b) and it is seen to be non-monotonic. 
Immediately after the laser pulse, the interface sets into motion (i). 
The maximum velocity $V_{max}$ occurs when a well formed jet just begins to appear as shown in image (ii), by which time the meniscus has lost the initial concave shape. This feature is observed irrespective of the jet speed, and is therefore compatible with the hypothesis of kinematic focusing due to the initial concave shape of the interface.
The subsequent deceleration (iii) is due to surface tension as suggested by the fact that the deceleration is suppressed at higher velocities, i.e., higher Weber numbers. %$We = \rho V^2_{max}D/\sigma \sim O(10^3)$ 
%when the effect of surface tension becomes less dominant than that of inertia.} 
As the jet becomes thinner, the rate of new surface generation decreases, surface tension becomes less important, and the jet tip velocity reaches an asymptotic value $V_j$; shortly thereafter a drop starts to form at the tip (iv) and it eventually detaches.
\\
It is worth mentioning that a very small fraction of the liquid receiving the initial impulse and focusing into the jet is in contact with the wall, and therefore its motion is not significantly affected by contact line phenomena. 
%These phenomena can play a minor role when the liquid remaining in the tube starts moving under the action of the growth and subsequent collapse of the vapor bubble. 
In any event, since the process is very rapid, such phenomena can only affect a thin liquid layer of the order of the viscous penetration length, which can be estimated as less than 5 $\mu$m. 
Most of the liquid is therefore essentially in an inertia-dominated, nearly inviscid motion and is not significantly affected by contact line phenomena. 
\\

\section{Parameter dependence}\label{sec:parameter_results}
In this section, we study the dependence of the jet speed on various control parameters. 
\subsection{Contact angle $\theta$}\label{subsection:angle}
\begin{figure*}
\centerline{\includegraphics[width=.75\textwidth]{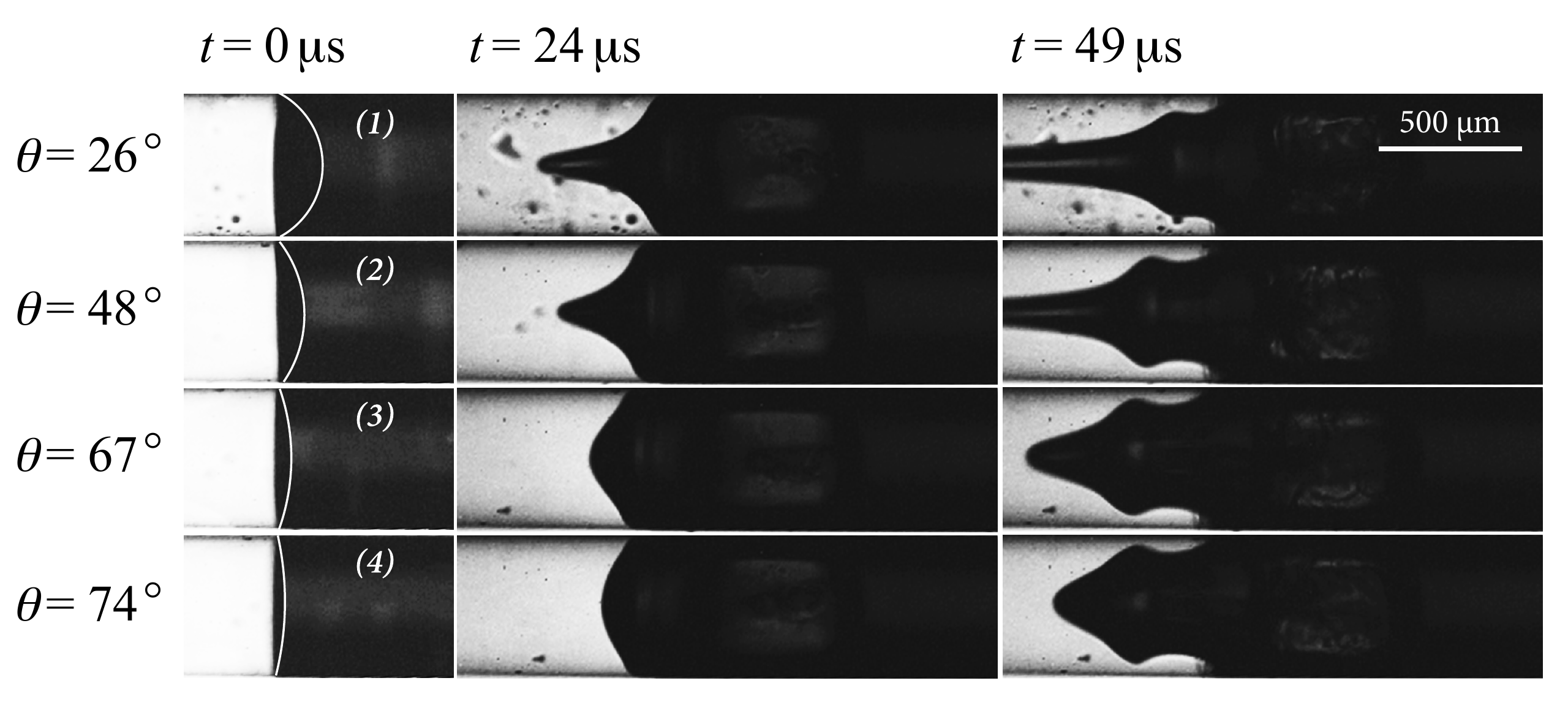}}
\caption{\label{fig:jet_shapes} Images showing the effect of varying the initial contact angle $\theta$ on the jet shape: (1) $\theta$ = $26^\circ$, (2) $\theta$ = $48^\circ$, (3) $\theta$ = 67$^\circ$ and (4) $\theta$ = 74$^\circ$. The first image shows the liquid in the capillary prior to the laser pulse, illustrating the initial contact angle and meniscus shape. The second image in each set is taken 24$\,\mu$s after the laser pulse, and the third image follows 25$\,\mu$s later. (The movies are available as supplemental material (1) for figure5-1.mov, (2) figure5-2.mov, (3) figure5-3.mov and (4) figure5-4.mov.)}
\end{figure*}

Since the concavity of the meniscus is crucial for the jet formation, the first parameter we investigated is the initial radius of curvature $R_0$ of the free surface. This quantity, which was determined by fitting a circle to the image of the free surface, is related to the contact angle $\theta$ of the liquid with the capillary by the formula
\begin{equation}
\label{eq:kappa}
R_0 = \frac{D}{2\cos{\theta}}.
\end{equation}
which is valid because the effect of gravity in this system is negligible as shown by the smallness of the Bond number $\Bo={\rho{g}{R_0^2}/{\gamma}}$ (where $\rho$ is the liquid density, $\gamma$ surface tension coefficient, and $g$ the acceleration of gravity) typically of the order of 10$^{-3}$. 
\\
The initial radius of curvature is varied by minor adjustments of the liquid volume by means of the syringe. For these tests, the laser focus was on the tube axis and the energy $E$ and the distance $H$ between focus position and meniscus are kept constant at $460\pm20\,\mu$J and $460\pm40\,\mu$m, respectively; experiments are carried on 500 $\mu$m diameter tubes. 
\\
Figure~\ref{fig:jet_shapes} shows four sequences taken with different initial contact angles, with three snapshots per sequence. The shape of the jet is significantly influenced by the radius of curvature. These results show that the more curved the meniscus shape (smaller radius of curvature and lower initial contact angle), the higher the jet velocity due to the increased focusing. For increasing initial contact angles, the jet becomes thicker and less focused. Eventually, when the initial contact angle is 90 degrees or larger, the focusing is lost, no jet is formed and the liquid is only pushed out as a plug by the expanding bubble. 
\\
These results confirm that the initial shape of the meniscus plays an important role for the acceleration of the jet as discussed in section~\ref{sec:results}.
For a quantitative analysis, we derive a scaling of the velocity increment due to the flow focusing as follows. 
\\
We assume that, due to the initial high pressure pulse, the liquid acquires a velocity $V_0$ normal to the initial free surface and the surface tension is negligible.
After a short time $\Delta t$, the free surface is still spherical with a smaller radius $R_0-V_0 \Delta t$.
The new velocity $V_0+\Delta V$ acquired due to this geometrical focusing can be estimated from the principle of mass conservation as
% At the moment of the shock wave impact, the initial velocity $V_0$ goes towards the center of the sphere with the initial curvature radius %of the interface %The initial shape of the meniscus, characterized by the curvature radius of the interface
\begin{equation}
\label{eq:continuous}
(V_0+\Delta V)(R_0-V_0 \Delta t)^2 = V_0 R_0^2.
\end{equation}
%where $\Delta V$ is the velocity increment. 
Neglecting a higher order term, %Taking the first order of the small increments and assuming  
we obtain the scale of the velocity increment as
\begin{equation}
\label{eq:acceleration}
\Delta V \sim \frac{V_0^2\Delta t}{R_0}.
\end{equation}
%by dimensional considerations.

Since the focusing effect is caused by the liquid particles approaching the
axis of the tube, it occurs on a characteristic length scale of the order
of the mean distance between the free surface of the liquid and the axis, which is
of the order of the tube diameter $D$. Thus the time scale $\Delta t$ for the
geometrical focusing can be estimated as
\begin{equation}
\label{eq:timescale}
\Delta t \sim \frac{D}{V_0}.
\end{equation}
This estimate is supported by the fact that, as described in \S~\ref{sec:results}, the
acceleration due to the focusing occurs approximately until the meniscus loses the concave shape.
Thus, upon combining eqs.~(\ref{eq:kappa}),~(\ref{eq:acceleration}), and~(\ref{eq:timescale}), the increase in velocity due to flow focusing can be estimated as

\begin{figure}
\centerline{\includegraphics[width=0.45\textwidth]{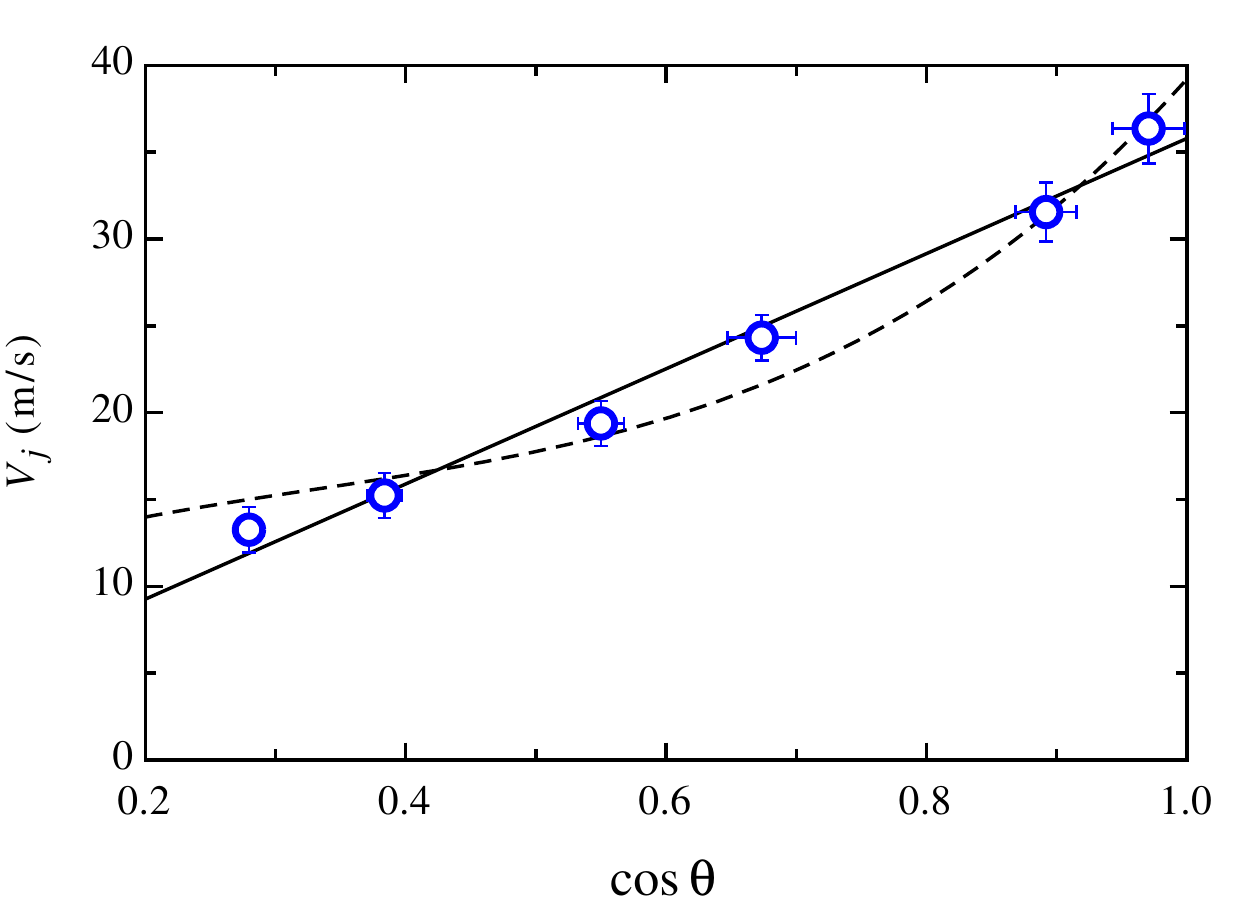}}
\caption{\label{fig:hydrophobic_plot} Asymptotic jet velocity $V_j$ as a function of the cosine of the initial contact angle. The solid line is a linear fit to the data.  
The dashed line is a $1-\sin{\theta}$ fit to the data suggested by \citet{Antkowiak2007}.} 
\end{figure}

%as described in \S~\ref{sec:results}, the acceleration due to the focusing occurs approximately until the meniscus loses the concave shape. In figure~\ref{fig:jet_shapes} the initial concave shapes vanish in a relatively short time even for the largest $\theta$. It is not expected if one takes $R_0$ as a length scale. This short time scale is more related to the overall size of the meniscus, which determines how fast the concave shape vanishes.
%%This is clearly due to the presence of the wall. 
%Therefore, the appropriate length scale here is the typical size of the meniscus, which scales with the tube diameter $D$.
 
\begin{equation}
\label{eq:focusedvelocity}
\Delta V \sim V_0\cos{\theta}.
\end{equation}
Thus, the velocity $V_j$ resulting from the focusing may be expected to be given by
\begin{equation}
\label{eq:velocity}
V_j \sim V_0 + \Delta V  = V_0 (1 + \beta\cos{\theta}),
\end{equation}
As shown in figure~\ref{fig:hydrophobic_plot}, this relation provides a reasonable fit to the data taken in the 500 $\mu$m tube by taking $\beta \simeq$ 13 and $V_0 \simeq$ 2.5 m/s. % for the present set of the data taken in .
%\textcolor{red}{The pressure of the shock wave is estimated as $\Delta p \sim 2V_0\rho c \approx$ 4 MPa}.
A more elaborated model will be described in~\citet{Peters2011}. 
The curve $1-\sin{\theta}$ suggested by~\citet{Antkowiak2007} is also shown in figure~\ref{fig:hydrophobic_plot}. 
Although, unlike the previous model which estimates the asymptotic jet velocity, the latter model is aimed to describe the initial velocity of the jet, its trend agrees with the present experimental data.
\\
If the initial velocity $V_0$ is due to the reflection of a shock wave from the free surface, the strength $\Delta p$ of the shock can be estimated as $\Delta p = 1/2 \rho c V_0$, where $c$ is the speed of sound in the liquid. 
With the previous value of $V_0$ we find $\Delta p \sim$ 20 atm.
\\
The initial velocity for the incompressible model can be estimated by assuming that it is acquired during a short time $\Delta \tau$ by the action of an over-pressure $\Delta p$. 
If the laser focus is at a distance $H$ from the free surface, this arguments leads us to 
$H \rho V_0 \sim \Delta p \Delta \tau$ \cite{Ory2000}. With $H$ = 460 $\mu$m, $V_0$ = 2.5 m/s, and $\Delta \tau$ = 6 ns, this estimate gives $\Delta p \sim$ 2000 atm, which is larger than the critical pressure of water, $\approx$ 218 atm. 
For higher velocity cases, the incompressible pressure estimate increases considerably \cite{Peters2011}.

\definecolor{ForestGreen}{rgb}{0,0.67,0}

\subsection{Distance $H$ between laser focus and free surface}\label{subsection:distance}
\begin{figure}
\centerline{\includegraphics[width=0.46\textwidth]{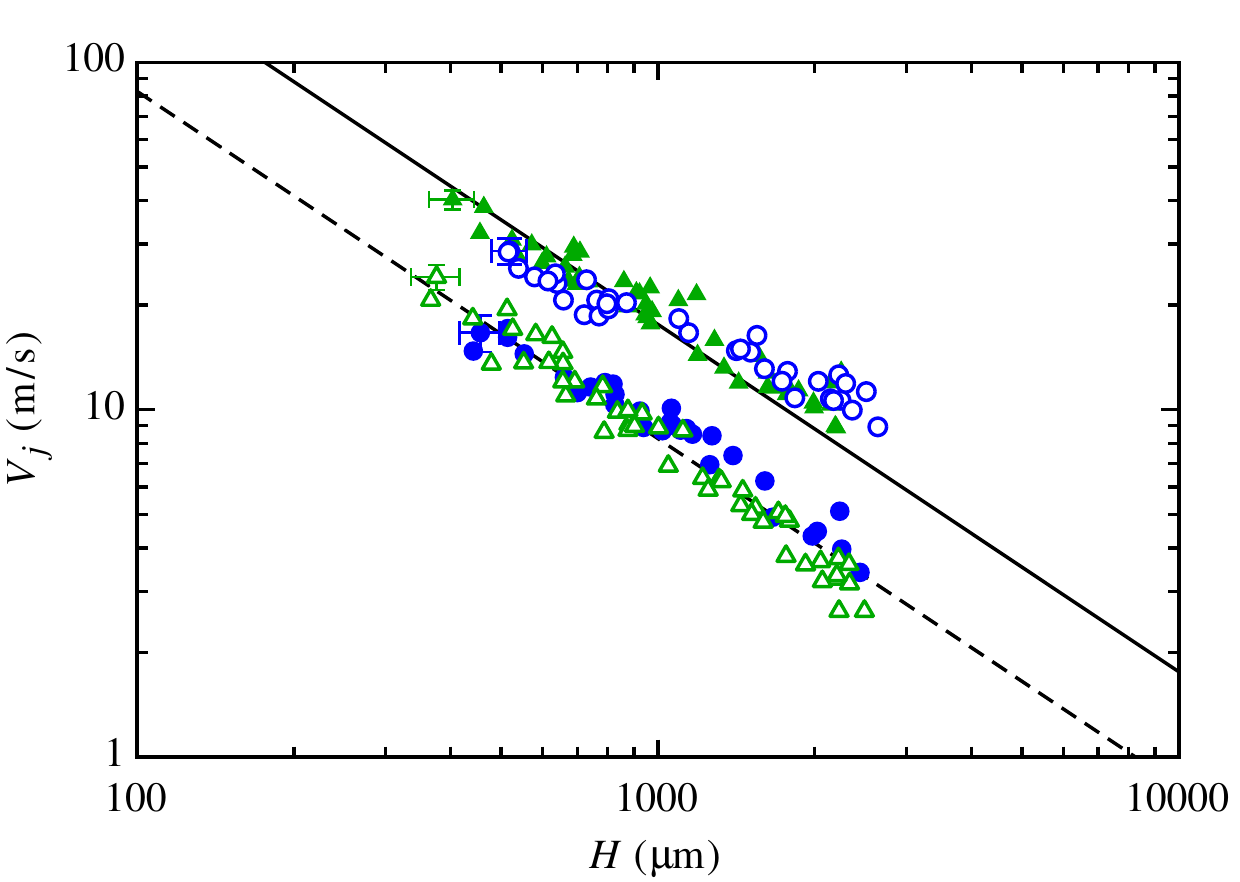}}
\caption{\label{fig:distance} Asymptotic jet velocity $V_j$ as a function of the distance $H$ between the laser spot and the free surface for different capillary diameters and energies. The triangles represent the data for the 200$\,\mu$m tube at $E=232\,\mu$J (\textcolor{ForestGreen}{$\blacktriangle$}) and $E=165\,\mu$J (\textcolor{ForestGreen}{$\vartriangle$}) and the circles show the results for the 500$\,\mu$m tube at $E=458\,\mu$J (\textcolor{blue}{$\Circle$}) and $E=305\,\mu$J (\textcolor{blue}{$\CIRCLE$}). The solid and dashed lines are showing a -1 power law. Typical error bars are shown for a few data points.}
\end{figure}

Figure~\ref{fig:distance} shows the experimental results for the asymptotic jet tip velocity $V_j$, defined in figure~\ref{fig:time_evolution}, as a function of the distance $H$ between the laser spot and the free surface (see figure~\ref{fig:full_setup}b). 
Two different capillary tubes with diameters of 200$\,\mu$m and 500$\,\mu$m, and two different energy levels $E$ for each, 165 $\mu$J and 232 $\mu$J, and 305 $\mu$J and 458 $\mu$J, respectively, were used. 
The data shows that, for both tube diameters, $V_j$ is inversely proportional to the distance $H$ over a decade. 
This dependence is particularly well satisfied at the lower energy levels. 
%%%%%%%%%%%%%%%%%%%%%%%%%%%%%%%%%
\\
It seems possible to rationalize this observation both on the basis of the incompressible flow focusing model and of the shock wave model.
In the framework of the former, one may note that the mass of the liquid slug set into motion is approximately proportional to $H$, and therefore so is its inertia.
In the framework of the shock wave mechanism, on the other hand, we recall that the pressure amplitude of a locally generated shock wave in a free fluid decreases inversely with the distance. 
In our conditions, the fluid is confined within the tube but the size of the initially vaporized liquid is at least one order of magnitude smaller than the tube diameter so that this geometrical attenuation may play a role. 
Another possibility is that there is viscous attenuation of the shock due to the no-slip condition at the tube wall (see e.g.~\cite{Mirshekari2009, Ngomo2010}).
%%%%%%%%%%%%%%%%%%%%%%%%%%%%%%%%%
Whatever the reason for the decreasing the shock pressure with increasing distance $H$, since the experiments show that the velocity is inversely proportional to the distance, this suggests that the velocity is proportional to the pressure. 
%Whatever the reason for the decreasing the shock pressure with increasing distance 
\\
\subsection{Absorbed energy $E$}\label{subsection:energy}
\begin{figure}
{\includegraphics[width=.45\textwidth]{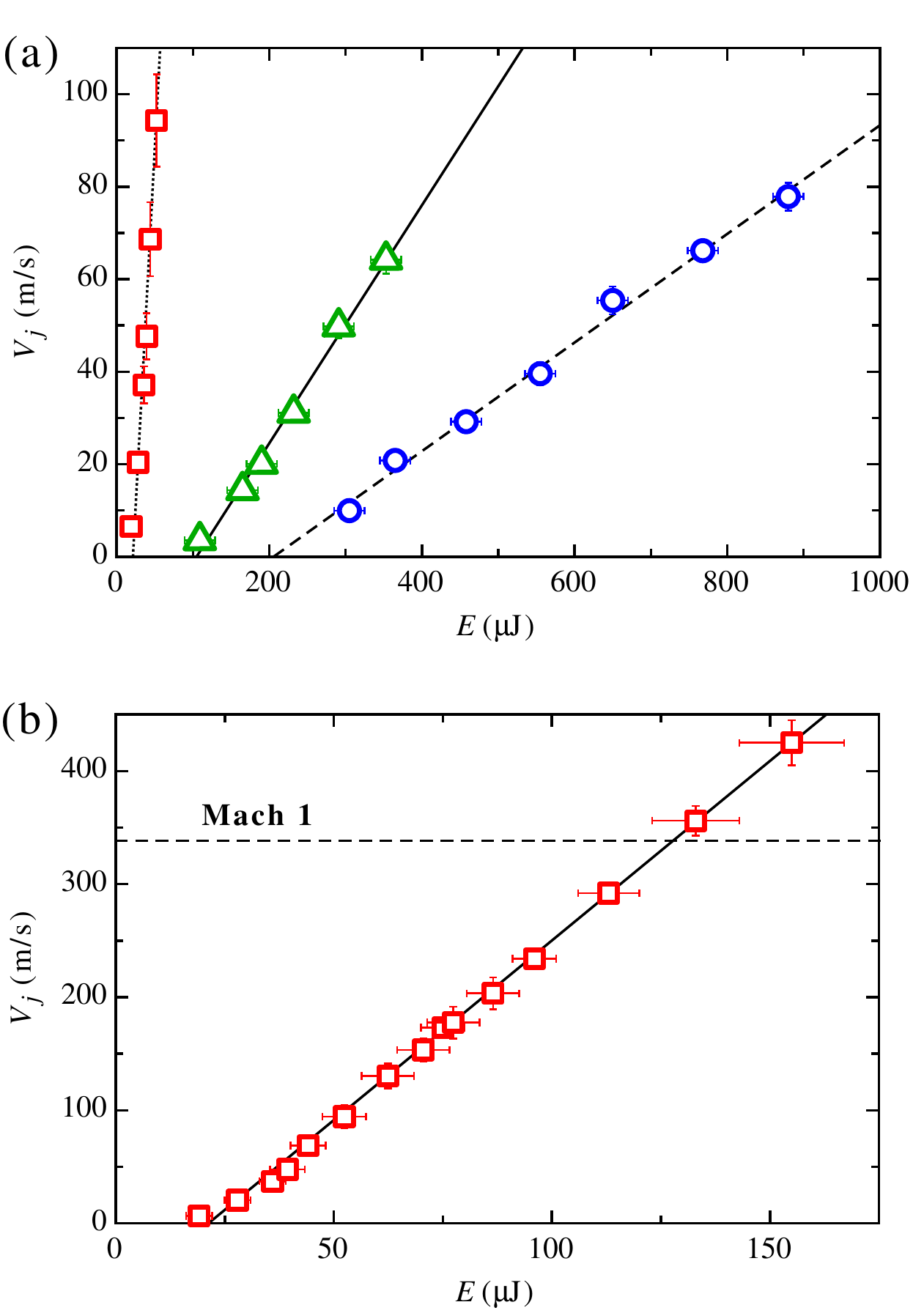}}
\caption{\label{fig:energy_all} (a) Asymptotic jet velocity $V_j$ as a function of the energy absorbed by the liquid in the capillary tubes. The squares (\textcolor{red}{$\square$}), triangles (\textcolor{ForestGreen}{$\vartriangle$}) and circles (\textcolor{blue}{$\Circle$}) represent the data for capillary tubes with 50$\,\mu$m, 200$\,\mu$m, and 500$\,\mu$m inner diameter, respectively. 
Each data point is the result of at least three measurements. The lines are linear fits to the data. 
The data for the 50$\,\mu$m diameter tube are shown on an expanded scale in (b). 
%(b) Asymptotic jet velocities $V_j$ for 50 $\mu$m of supersonic microjets as a function of the energy absorbed by the liquid in the capillary tubes with 50$\,\mu$m inner diameter. 
Each data point is the result of at least five measurements. The linear dependence on the energy is seen to hold also for supersonic speeds.}
\end{figure}
The next parameter of interest is the energy absorbed by the liquid.
For this part of the study, the laser focus is on the tube axis and its distance from the meniscus was kept constant at $410\pm40\,\mu$m, $390\pm40\,\mu$m, and $600\pm40\,\mu$m for the 50$\,\mu$m, 200$\,\mu$m, and 500$\,\mu$m diameter tubes, respectively.
\\
Figure~\ref{fig:energy_all}(a) shows the asymptotic jet velocity $V_{j}$ vs. the absorbed energy for different tube diameters. 
The data are well fitted by a linear function with a positive intercept at $V_j = 0$. This feature embodies the existence of a threshold $E_{heat}$ below which no jet is formed.
The threshold values for the 50$\,\mu$m, 200$\,\mu$m, and 500$\,\mu$m capillary tubes are approximately 20$\,\mu$J, 100$\,\mu$J, and 200$\,\mu$J, respectively.
Thus, the threshold value is an increasing function of the tube diameter $D$. 
It is interesting to note that, for the 50$\,\mu$m capillary tube, the linear relation between $V_j$ and $E-E_{heat}$ is preserved even when the jet speed becomes supersonic as shown in figure~\ref{fig:energy_all} (b). 
The absorbed energy for all cases is of the order of 10$^{9}$ J/m$^3$, which is of the same order as the vaporization enthalpy of water at normal conditions.
\\
As noted before, in the framework of the shock wave model, the experimental results of \S~\ref{subsection:distance} suggest that the velocity is proportional to the pressure.
This linear relation would then imply that the pressure of the shock wave at the meniscus is proportional to the absorbed energy.
From the incompressible viewpoint, one may conclude that the initial impulse imparted to the liquid is proportional to the absorbed energy.
\\
The slopes of the data for the 50$\,\mu$m, 200$\,\mu$m, and 500$\,\mu$m diameter tubes are 3.09, 0.26, and 0.12 m/(s$\cdot\mu$J), respectively. 
Thus, for the same absorbed energy, the jet speed decreases with increasing tube diameter. 
The effects of the diameter $D$ on the jet velocity are discussed further in \S~\ref{subsection:diameter}.
\\
In figure~\ref{fig:energy_all}(b) no data points at higher $V_j$ could be acquired because in the current configuration the glass tube breaks at the point where the laser focuses when the energy $E$ is too high. 
\subsection{Diameter of microcapillary $D$}\label{subsection:diameter}
\begin{figure}
\centerline{\includegraphics[width=0.45\textwidth]{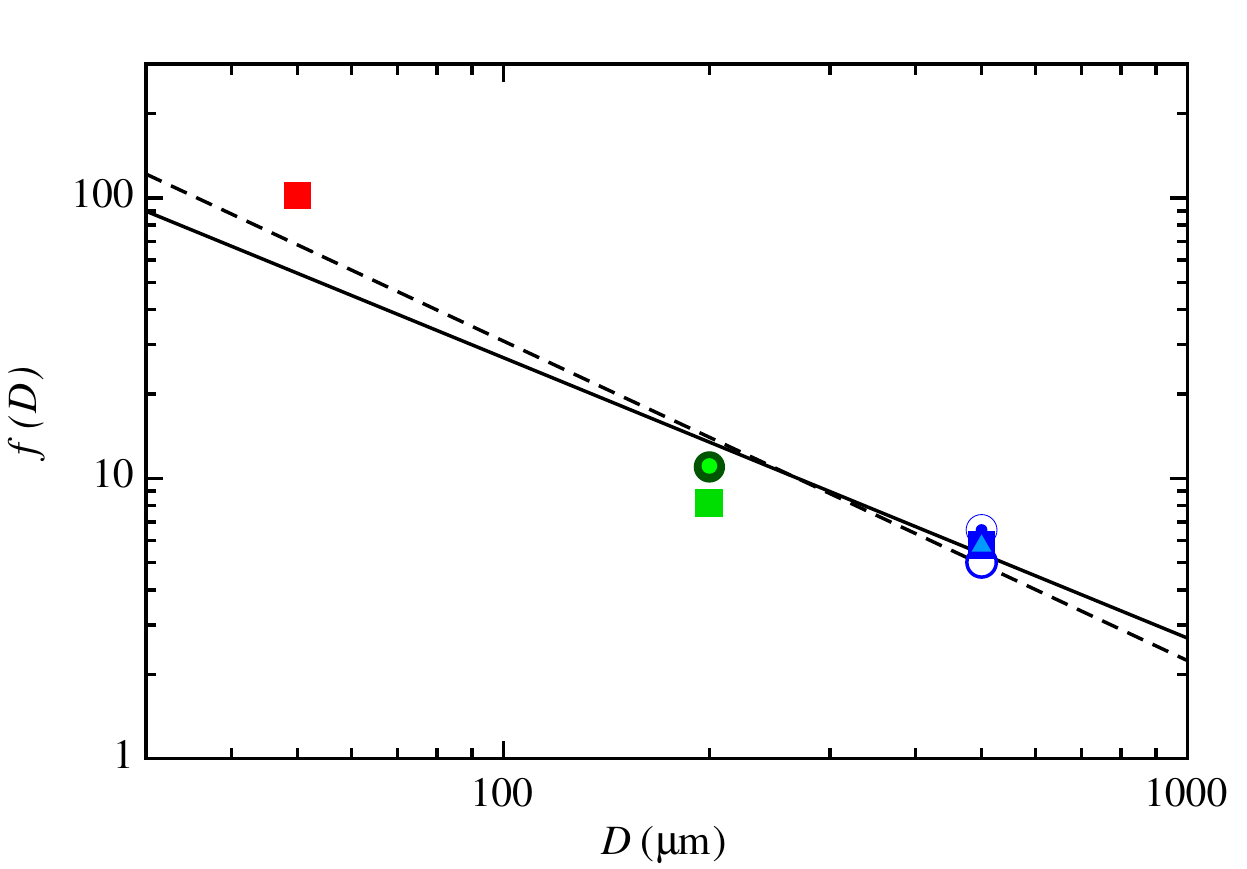}}
\caption{\label{fig:diameter_fit} Pre-factor $f(D)$ from equation~(\ref{eq:prefactor}) versus the capillary diameter. The circles, squares, and diamonds are constants obtained from the energy dependence, distance dependence, and initial contact angle dependence experiments, respectively. The value of the pre-factor increases with smaller capillaries.  A best fit power law (dashed line) with power -1.14 is shown as well as a line with power -1 as shown in equation~(\ref{eq:empirical}) (solid line), which shows less agreement with the data, in particular for small $D$.}
\end{figure}
In the study of the dependence of the jet velocity on the tube diameter, it was necessary to adjust the energy ranges depending on the diameter. On the basis of the results reported above, it has been concluded that, to a good accuracy, $V_j$ is proportional to ($E-E_{heat}$) and ($1+\beta\cos\theta$) and inversely proportional to $H$.
We express this dependency by writing
\begin{equation}
\label{eq:prefactor}
V_j \simeq f(D)\frac{(E-E_{heat})(1 + \beta \cos{\theta})}{H},
\end{equation}
where the pre-factor $f(D)$ embodies the dependence on the diameter. By trying a power law dependence, $f(D) = f_0D^{-\alpha}$, the data is best fitted by $\alpha \simeq$ -1.14, which may be considered close to -1.
\\
Especially the data for the 200$\,\mu$m and 500$\,\mu$m tubes agrees well with the line of slope -1, while that for 50$\,\mu$m is above the line with slope -1. 
This could be attributed to the fact that different effects, such as laser focusing, start to play an important role at smaller tube sizes due to the higher curvature of the tube. 
According to the impulse pressure description suggested by \citet{Antkowiak2007} and the analysis leading to eq. (\ref{eq:focusedvelocity}), the diameter of the tube should not play a role.
A possible reason for this dependence is that the same energy delivered to a smaller diameter tube will give rise to a faster jet due to a smaller inertia. 
To strengthen this hypothesis it may be noted that, if the kinetic energy is constant for a given absorbed energy $E$, since the mass of the accelerated liquid is approximately proportional to $D^2$, it follows that the velocity would decrease proportionally to $D^{-1}$. 
From the results that were presented in the previous sections, the following empirical law is then derived:
\begin{equation}
\label{eq:empirical}
V_j \simeq C_0 \frac{(E-E_{heat})(1 + \beta \cos{\theta})}{H D}.
\end{equation}
Calculating $C_0$ by fitting the data in figure~\ref{fig:diameter_fit} we get $C_0=0.0027\,(\text{Pa}\cdot\text{s})^{-1}$. 
It is interesting to note that $C_0$ has the the dimensions of an inverse dynamic viscosity. This feature appears to be accidental as viscous effects are not expected to play a significant role as noted before.
\\
\subsection{Focus offsets, $l_{v}$ and $l_{h}$}\label{subsection:focus}
 
\begin{figure}
\centerline{\includegraphics[width=.48\textwidth]{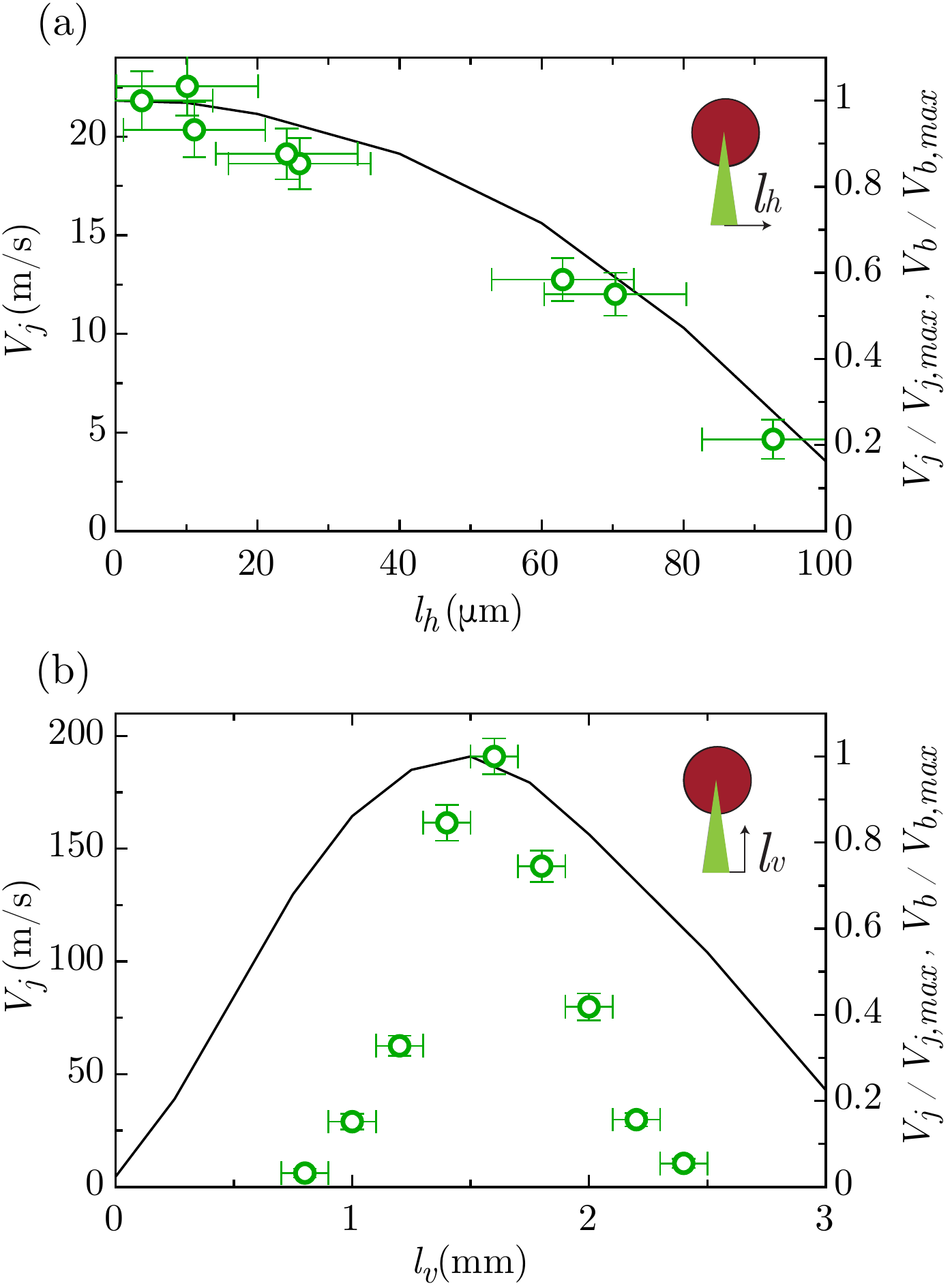}}
\caption{\label{FocusOffset} (a) Asymptotic jet velocity $V_j$ (\textcolor{ForestGreen}{$\Circle$}) as a function of the horizontal focus displacement of the laser for the 200$\,\mu$m tube. % with $E$ \textcolor{red}{XXXX} and $H$ \textcolor{red}{XXXX}. 
The black line shows the normalized vaporized liquid volume determined by geometrical optics approximation in both graphs (right axis). (b) Asymptotic jet velocity $V_j$ as a function of the vertical focus displacement of the laser. %for  $E$ \textcolor{red}{XXXX} and $H$ \textcolor{red}{XXXX}. 
The inserts show the capillary and the laser focus (triangle) and the directions of $l_h$ and $l_v$ as seen from the capillary opening.}
\end{figure}
 
\begin{figure}
\centerline{\includegraphics[width=.5\textwidth]{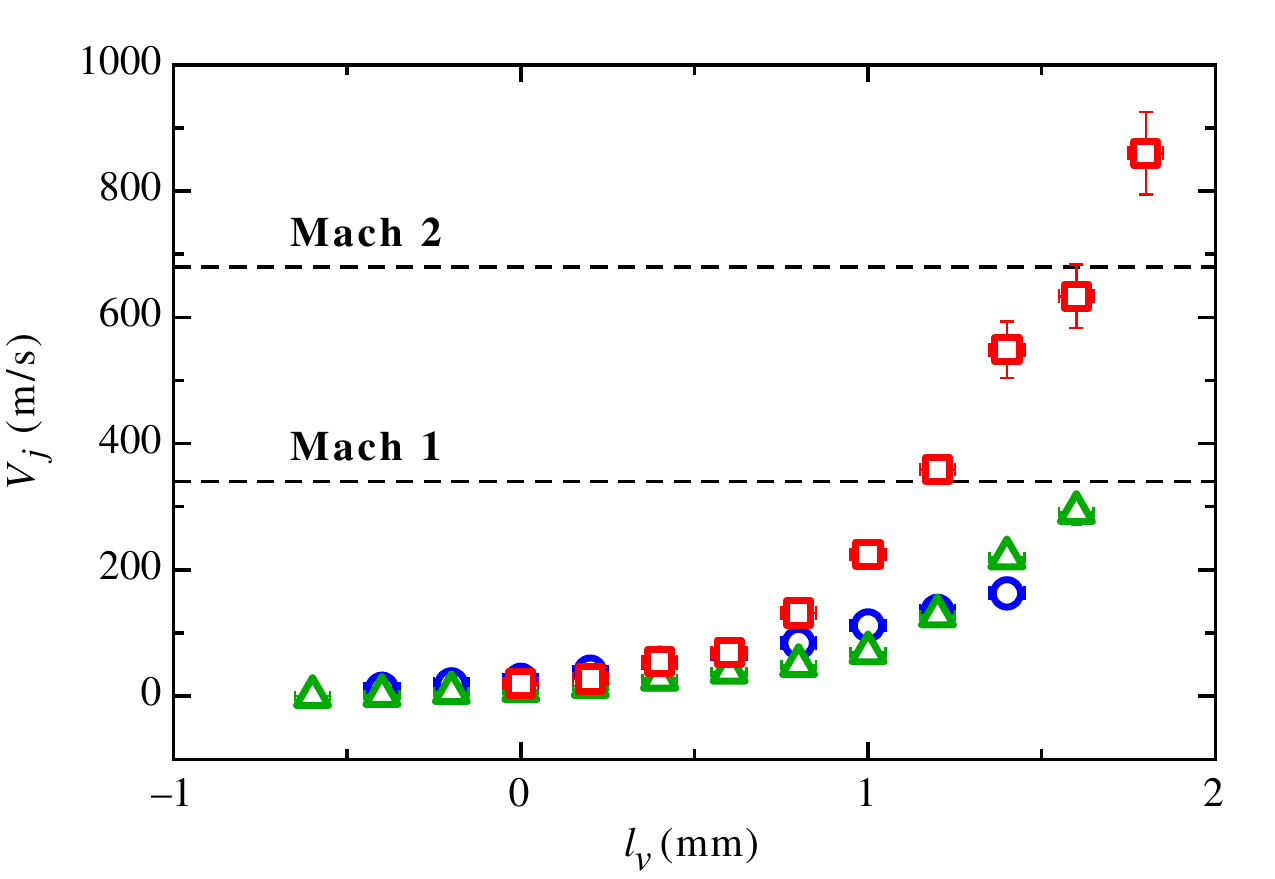}}
\caption{\label{fig:focus_variation_total} Asymptotic jet velocity $V_j$ as a function of the focus offset of the laser for different capillary diameters. The circles (\textcolor{blue}{$\Circle$}) represent measurements for the 500$\,\mu$m tube; the triangles (\textcolor{ForestGreen}{$\vartriangle$}) refer to 200$\,\mu$m tube; the squares (\textcolor{red}{$\square$}) are for the 50$\,\mu$m. Each data point is the result of at least three measurements.}
\end{figure}
The sensitivity of the jet to the laser focus position relatively to the capillary axis was studied by displacing it in the horizontal ($l_h$) and vertical directions ($l_v$) (see figure \ref{fig:full_setup}b). The results are shown in figure \ref{FocusOffset}.
 \\
Figure \ref{FocusOffset}(a) indicates that the jet velocity decreases monotonically for increasing $l_h$ all other conditions being held constant. We hypothesize that this trend is explained by a reduced volume of the vaporized liquid. To test this hypothesis, the jet velocity is compared to the initial laser-induced vaporized volume as estimated with the geometrical optics approximation described in Appendix A. As shown in figure \ref{FocusOffset}(a), the jet velocity roughly scales with the calculated volume of the liquid vaporized by the laser pulse.
\\
Figure \ref{FocusOffset}(b) shows the jet speed versus the distance of the focus from the axis of the tube $l_v$. The jet velocity increases with increasing $l_v$ up to $l_v \approx 1.5$ mm, and then decreases for larger $l_v$. 
The rise is likely caused by the larger area of the capillary tube surface illuminated by the laser, which increases as $l_v$ is increased by moving the laser towards the tube. As the vaporization always occurs at the capillary wall, a larger vapor mass will be created by increasing $l_v$. However, when the laser is moved too close to the capillary, the energy per unit area of the laser beam decreases and eventually drops below the threshold required for bubble formation. This situation was also modeled by geometrical optics. As shown in figure \ref{FocusOffset}(b), the maximum jet speed and maximum bubble size are found at the same offset. Even though the overall trend is the same, the normalized values of the vaporized volume deviate from those of the jet velocity.
This difference is not surprising since there is no reason to think that vaporized volume and jet growth are linearly related to each other. The analysis of these aspects is beyond the scope of this work; they will be studied in future investigations.
\\
The velocity increase with increasing $l_v$ shown in figure \ref{FocusOffset}(b) is found also for other tube diameters
as shown in figure~\ref{fig:focus_variation_total}, where the jet speed versus the vertical focus offset $l_v$ is plotted for different diameters down to 50$\,\mu$m. Jets with a velocity up to 850 m/s could be consistently produced with the smallest capillaries. 
Measurements beyond this velocity could not be obtained, as the gradual increase of the absorbing liquid volume with increasing $l_v$ eventually leads to such a violent bubble expansion that the capillary breaks.
We also observed that, at the higher energies, the wall shear stresses exerted by the liquid pushed out of the capillary by the expanding bubble after the thin jet formation is so large as to shear off the capillary tip. It is possible that, with different materials, the velocity of the jet can be increased. When the jet velocity approaches the speed of sound in water, some limitations to further increases may arise due to the dominance of compressibility effects.
%\textcolor{red}{Although the shock wave requires compressibility, whether the generation of the microjet needs compressibility or not is still an open question.}
\\
\section{Summary and conclusions}
\label{sec:conclusions}
The dynamics of the high-speed microjet generated by a laser-induced
rapid vaporization of water in a microtube has been studied. It has
been shown that the jets so generated can reach speeds as high
as 850 m/s with good controllability.
\\
The dependence of the jet velocity upon various controlling
parameters has been investigated in a series of experiments the
results of which have been summarised in the empirical relationship
provided in equation~(\ref{eq:empirical}). This equation shows the
effect of the distance between the laser focus and the liquid meniscus
at the mouth of the micro-tube, the absorbed laser energy, the
liquid-tube initial contact angle and the tube diameter.
\\
The jet velocity exhibits an inverse proportionality to the tube diameter and
to the distance of the laser focus from the free surface, while it is
proportional to the absorbed energy above a threshold value.
The velocity is very strongly dependent on the curvature of the free surface,
which is a function of the tube-liquid initial contact angle. 
It is also critically dependent on the amount of liquid vaporized and on its distance from
the free surface, both of which can be varied by varying the position
of the laser focus.
In particular, it has been found that the offset of the laser focus with respect to the tube axis has a strong and
very non-trivial effect. 
To elucidate the origin of this result we have
used a geomterical-optics construction coupled with the Beer-Lambert law
to determine approximately the size of the region where vaporization
occurs and the absorbed energy. While still preliminary, the results of this
analysis are in general agreement with the data.  
Further theoretical and numerical investigations of the jetting phenomenon will be addressed
in~\citet{Peters2011}.
\\
A priori results of the paper can be interpreted according to two different scenarios, one in which the shock wave generated by the nearly instantaneous vaporization plays a dominant role, and one in which the phenomenon is essentially incompressible. % and dominated by the kinematic focusing due to the initial shape of the free surface.
The estimate of the overpressure provided for the relatively low-velocity case in \S ~\ref{subsection:energy} suggests that, according to the incompressible model, a much higher pressure is necessary to account for the observations as compared to the shock wave model.
For higher velocity cases, the incompressible overpressure becomes even much larger than the critical pressure of water \cite{Peters2011}.
These considerations lend support to the shock wave interpretation of the results presented in this paper.
\\
The insights gained through this research and the ability to generate focused, controllable, and high velocity microjets open new doors for the realization, among others, of reliable needle-free drug delivery systems.
\\
\section*{Acknowledgments}
We thank Rory Dijkink for starting the experiments. We also thank Christophe Clanet, Alexander Klein, and 
Gerben Morsink for helpful discussions. We gratefully acknowledge the support of this work by FOM (Stichting voor Fundamenteel Onderzoek der Materie). 

\vspace*{1cm}

\begin{appendix}
\section{Estimate of the liquid energy absorption by geometrical optics}
A geometrical optics approximation was developed to analyze energy absorption of the laser beam energy absorption by the liquid. The three main components of this calculation are: (1) splitting up the Gaussian laser beam into many different rays each one with a representative energy; (2) following the path of each ray through the capillary, including (full or partial) wall-reflections; (3) using the Beer-Lambert law to model the local energy absorbed by the liquid. This law describes the local irradiance $I$, and is given by $I = I_0 \cdot 10^{-\epsilon s}$, with $I_0$ the beam irradiance at the cylinder outer surface of the tube, $\epsilon$ the absorption coefficient, $s$ the arc length along the partial of the ray propagating in the liquid. The absorption coefficient was measured as $\epsilon = 84 \cdot 10^3\mathrm{m}^{-1}$; the local energy loss of a ray equals $\partial I /\partial s $.
\\
The liquid volumes are discretized into cells constituting a grid around the central loss area. The local energy lost by each ray in each cell was calculated from the Beer-Lambert law and add to the liquid in the cell. Subsequently, for each cell, the total energy absorbed over the duration of the light pulse was compared with an estimate of the energy necessary for evaporation at room conditions, namely $E_{boil} = \rho (C_p \Delta T + \Delta H_{vap})\Delta V_c$, with $\rho$ the liquid density, $\Delta T = 80$ K the difference between the initial liquid temperature and the boiling temperature, $C_p = 4181$ J/(kgK) the heat capacity at constant pressure, $ \Delta H_{vap} = 2.26$ MJ/kg the vaporization enthalpy, and $\Delta V_c$ the cell volume. If the energy absorbed by a cell exceeded $E_{boil}$, the cell volume was assumed to be vaporized instantaneously. Summation of the vaporized-cell volumes then provided a measure for the initial bubble volume and position.

\end{appendix}

\bibliographystyle{apsrev4-1}

\end{document}